%% file: main.tex
\documentclass[cameraready]{Interspeech}

\title{HALAS: A Human-Annotated Dataset of Hallucinations of Modern ASR Systems}

\author[orcid=0009-0004-3021-7814]{Mateusz}{Barański}
\author[orcid=0000-0002-3897-181X, correspondingauthor]{Jan}{Jasiński}
\author[orcid=0000-0002-6708-876X]{Julitta}{Bartolewska}
\author[orcid=0000-0003-2921-6502]{Marcin}{Witkowski}
\author[orcid=0000-0002-7834-6920]{\\Konrad}{Kowalczyk}

\address{Signal Processing Group, Institute of Electronics, AGH University of Krakow, Poland}

\email{mbaranski@agh.edu.pl, jjasinsk@agh.edu.pl, bartolew@agh.edu.pl, witkow@agh.edu.pl, konrad.kowalczyk@agh.edu.pl}

\keywords{automatic speech recognition, speech-to-text hallucinations, error detection, human-annotated dataset, ASR error analysis}

\usepackage{comment}
\usepackage{url}

\begin{document}

\maketitle
\begin{abstract}

End-to-end Automatic Speech Recognition (ASR) systems hallucinate on natural speech, yet existing mitigation methods are typically evaluated on non-speech or artificially corrupted audio. We introduce HALAS, the first human-annotated dataset of naturally occurring hallucinations from seven state-of-the-art ASR models on real unprocessed earnings call recordings. HALAS provides span-level labels, enabling analysis of hallucination patterns and their severity. Our analysis reveals strong cross-model vocabulary overlap and confirms that hallucinations also occur for almost correctly transcribed speech (characterized by a low Word Error Rate). The proposed benchmark with HALAS shows that the character and semantic-level metrics used as a proxy for hallucination detection reach 81\% ROC-AUC, while state-of-the-art detection methods achieve an F1 score of only 53.1\%. As such, HALAS establishes the first rigorous non-artificial benchmark for the detection and mitigation of ASR hallucinations.

\end{abstract}

\input{intro_mw}

\section{Dataset Creation Methodology}
\label{sec:methodology}

\begin{figure}[!tb]
\vspace{-10pt}
\begin{minipage}[b]{1.0\linewidth}
  \centering
  \centerline{\includegraphics[width=8cm]{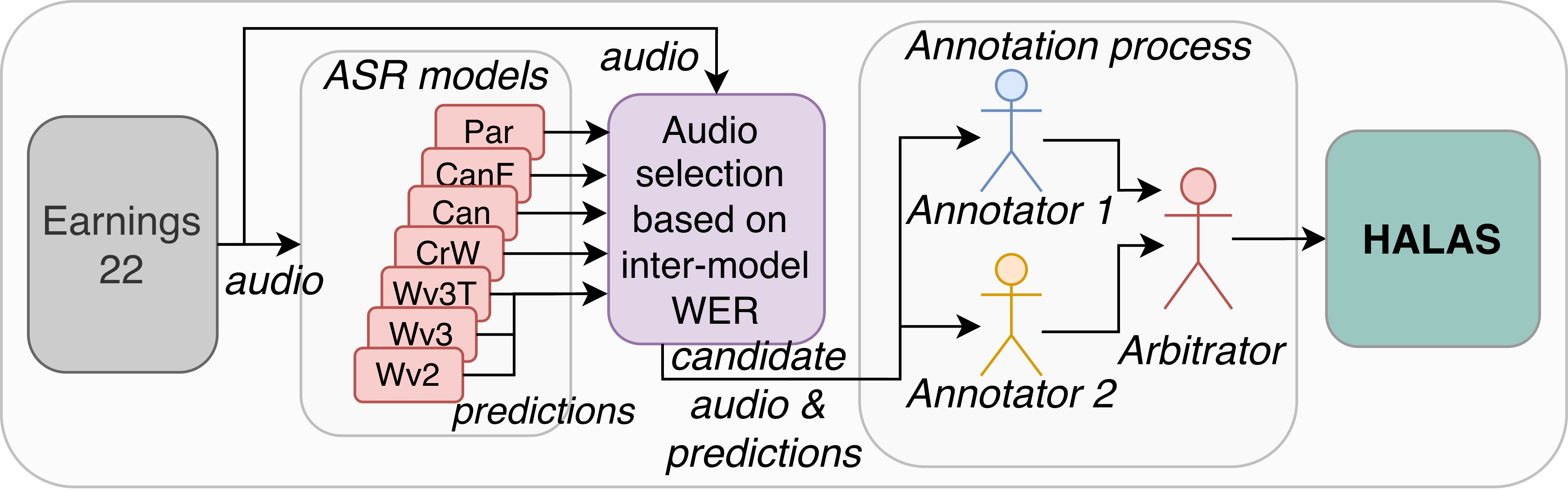}}
\end{minipage}
\caption{Diagram of audio file pipeline in HALAS creation.}
\label{fig:AnnotationDiagram}
\vspace{-18pt}
\end{figure}

The following section presents the entire dataset creation process, depicted in Fig. \ref{fig:AnnotationDiagram}.
We assumed that real-world spontaneous speech recordings of varying lengths and qualities would pose challenges to ASR models, thus increasing the chances of hallucinations. 
We used the Earnings 22 (E22) dataset~\cite{earnings22} as it contains earnings calls in English conducted by speakers from 27 countries. The provided audio files\footnote{\href{https://huggingface.co/datasets/distil-whisper/earnings22}{\url{https://huggingface.co/datasets/distil-whisper/earnings22}}} with total length 119~h are divided into 57390 segments with the 5th and 95th percentiles of duration equal to 0.6 s and 17.6 s, respectively.

We selected 7 state-of-the-art ASR models that achieve low WER results across various datasets in the Open ASR Leaderboard \cite{srivastav2025openasrleaderboardreproducible} (as of June 1, 2025). The selected models include 4 versions of OpenAI's Whisper~\cite{radford2023robust}: large v3 (Wv3), large v3 Turbo (Wv3T), large v2 (Wv2), and Crisper Whisper (CrW)~\cite{wagner2024crisperwhisperaccuratetimestampsverbatim} (a modified fine-tuned version of Wv2), as well as 2 autoregressive encoder-decoder (AED) models: Nvidia Nemo Canary-1B (Can)~\cite{puvvada2024moreaccuratespeechrecognition} and its parameter-reduced version Canary-1B-Flash (CanF)~\cite{zelasko2025traininginferenceefficiencyencoderdecoder}; along with another model from Nvidia Nemo, Parakeet-TDT version 2 (Par)~\cite{xu2023efficientsequencetransductionjointly}. 
We used Hugging Face implementations for CrW, Wv2 and GitHub versions for the remaining models. To our knowledge, all of these models did not utilize the E22 dataset during training. Inferences were conducted using all 7 models with the default parameters proposed by their creators on the entire E22 dataset.

\subsection{Speech data and state-of-the-art ASR models}
\label{sec:methodology_asr}

To maximize hallucination prevalence in our annotation pool, we used inter-model disagreement, namely inter-model WER, as a proxy for audio difficulty. The intuition is that when the same audio yields substantially different transcriptions across models, it is likely to be genuinely challenging and prone to severe recognition errors, including hallucinations. The following candidate selection procedure was utilized:
\begin{itemize}
    \item \textbf{Preprocessing:} All files marked in E22 with the tags "inaudible" and "foreign language" were discarded.
    \item \textbf{Candidate Selection:} First, to prevent repetitive looping from dominating the disagreement metric, we removed repetitions of the same words or phrases and then, for fair comparison, we also normalized all predictions%
    \footnote{ BasicTextNormalizer from \url{https://pypi.org/project/whisper-normalizer} was used.}.
    Next, we calculated the average inter-model WER across all possible pairs of ASR models (using one model's prediction as the reference for another) for each audio file. The candidates for the annotation process were selected from the top of the list of segments with descending average inter-model WER values. Note that looping removal and normalization were applied solely for metric calculation, thus in the next annotation steps, original predictions of selected audio files were used.
\end{itemize}

\subsection{Hallucination Annotation process}
\label{sec:methodology_anotation}

HALAS was annotated by 10 paid professional annotators (English B2+). In line with hallucination definitions in the literature \cite{frieske2024hallucinations,ji2023survey, atwany2025lost}, the instruction for annotators defined hallucinations as "\textit{A prediction, or a fragment thereof, that has no phonetic correspondence with the content of the audio signal analyzed by the ASR model}". The distinction between hallucinations and phonetic errors was explained with specific examples. Marking was to be conducted solely on the basis of the comparison of each prediction with the audio recording, specifically not the reference text provided by the dataset. 
For each audio sample, an annotator was asked to perform the following steps: 
\begin{itemize}
    \item \textbf{Audio Validation:} The audio was played and carefully analyzed, allowing the annotator to identify and mark any issues, such as the absence of speech, simultaneous speech from multiple speakers, or predominantly non-English speech. Such files were excluded to avoid annotation ambiguity. %
    \item \textbf{Prediction Hallucination Annotation:} For each ASR model, the annotators were asked to read the prediction, listen to the audio recording once more, and mark the words or phrases that follow the hallucination definition and tag them as "\textit{Hallucination}," "\textit{Looping}," or "\textit{Looping Hallucination}". The distinction between the last two tags is based on whether the word or phrase, which is repeated erroneously, was present in the speech or was itself a hallucination.
\end{itemize} 

\input{Tables/tab_HalasSplits}

As depicted in Fig.~\ref{fig:AnnotationDiagram}, each file was processed by 2 independent annotators, with a third annotator acting as a decisive arbitrator. The initial annotation agreement was high (Cohen's kappa = 0.87)~\cite{cohen1960coefficient}. Although Earnings-22 provides reference transcripts, annotators frequently flagged them as incorrect. To ensure reliable ground truth, the arbitrator re-checked and corrected all references, marking 14\% of examples as inaudible where the spoken content could not be clearly discerned.

Note that HALAS is intentionally constructed to contain a high proportion of hallucinations by sampling utterances exhibiting inter-model disagreement. As a result, the dataset does not reflect the frequency of real-world hallucinations in typical ASR deployment scenarios. Nevertheless, the annotated hallucinations originate from real speech recordings and therefore capture naturally occurring error patterns of ASRs.

\input{Tables/tab1}

\section{Hallucination Dataset}
\label{sec:dataset}

The HALAS dataset contains predictions with hallucination annotations for 3,611 audio files from the Earnings 22 dataset and 7 state-of-the-art ASR models listed in Section \ref{sec:methodology}. The dataset is provided as a CSV file, where each row corresponds to a single audio sample identified by its E22 filename (in format \textit{SEGMENT-ID\_FILE-ID.wav}). Each row encompasses the predictions and corresponding human annotations for all seven ASR models. Each model has an utterance-level hallucination label and annotations detailing the text and letter span of the hallucination or looping. \textit{Train} and \textit{test} splits are provided in which files were partitioned by source meeting and stratified by average WER, hallucination rate and duration. To ensure a stable evaluation benchmark, the \textit{test} set was filtered to include only audio exceeding 1.0 second with at least three words. This yielded a reliable \textit{test} split with a hallucination rate of 22.6\%, compared to 33.6\% in the \textit{train} split. The summary of the splits of the HALAS dataset are presented in Table \ref{tab:benchmarks_splits}. The dataset together with all supplementary material, including commit hashes and inference parameters for all models, is available on the GitHub repository\footnote{\href{https://github.com/DSP-AGH/HALAS/tree/main}{\url{https://github.com/DSP-AGH/HALAS/tree/main}}} and HuggingFace\footnote{\href{https://huggingface.co/datasets/MatBar99/HALAS}{\url{https://huggingface.co/datasets/MatBar99/HALAS}}}.

\subsection{Quantitative and qualitative analysis of the dataset}

Table \ref{tab:dataset} presents the number of audio samples with annotated errors (marked as hallucination or looping) for each ASR model, along with utterance level hallucination and looping rates. Hallucination rate varies between 21.4\% and 43.8\%, confirming that all investigated models hallucinate on HALAS audio data. Looping occurs only in 1.1\% cases on average. The values ranging from 31.46\% to 110.5\% show the difficulty of HALAS audio.

\begin{figure}[!t]
\vspace{-10pt}
\centering
  \centerline{
  \includegraphics[width=8.5cm]
  {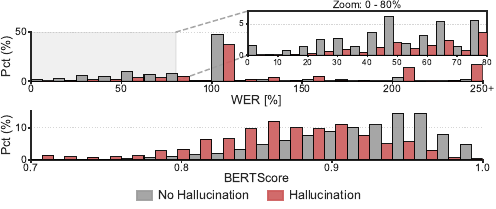}
  }
\vspace{-8pt}
\caption{All model WER (top) and BERTScore (bottom) distribution. Extreme WER aggregated in 250+ overflow bin.}
\vspace{-17pt}
\label{fig:werhistogram}
\end{figure}

Next, we assess the hallucination distributions in the dataset similarly to \cite{baranski2025investigationwhisperasrhallucinations}. To this end, for each model, all marked hallucinations were normalized, added to a single list, and the number of appearances of each of them was calculated, giving us a distribution of hallucination texts. Analyzing these results, we see that for each model, the distribution is heavily skewed toward a small set of the most frequent phrases. On average, 55\% of marked hallucinations correspond to the top 10 phrases of a given model. Expanding to the top 30 phrases increases this coverage to 75\%. 
This pattern is consistent with findings from studies that examined Whisper's hallucinations in non-speech audio~\cite{baranski2025investigationwhisperasrhallucinations}. Our results indicate that this phenomenon is observed across all tested models, with it being less pronounced for Can and CanF, while Par results are even more heavily skewed. 

Figure \ref{fig:werhistogram} shows the distributions of (per utterance) WER and BERTScore \cite{bert}. These metrics are chosen to analyze structural and semantic differences. The WER histogram demonstrates a large overlap between clean and hallucinated utterances. This indicates that this metric cannot reliably isolate hallucinations unless the erroneous insertions are catastrophically long, as seen in the 200\% and the 250\%+ overflow bins. Peaks at multiples of 100\% are an artifact of the many short audio segments, with few words in their references. Interestingly, hallucinations also occur for low values of WER reaching 6.25\%. While the BERTScore distribution shows a visible leftward shift for hallucinations, there remains substantial overlap with non-hallucinated text in the 0.80 to 0.95 range.

\begin{figure}[!tb]
\vspace{-10pt}
\begin{minipage}[b]{1.0\linewidth}
  \centering
  \centerline{\includegraphics[width=6.5cm]{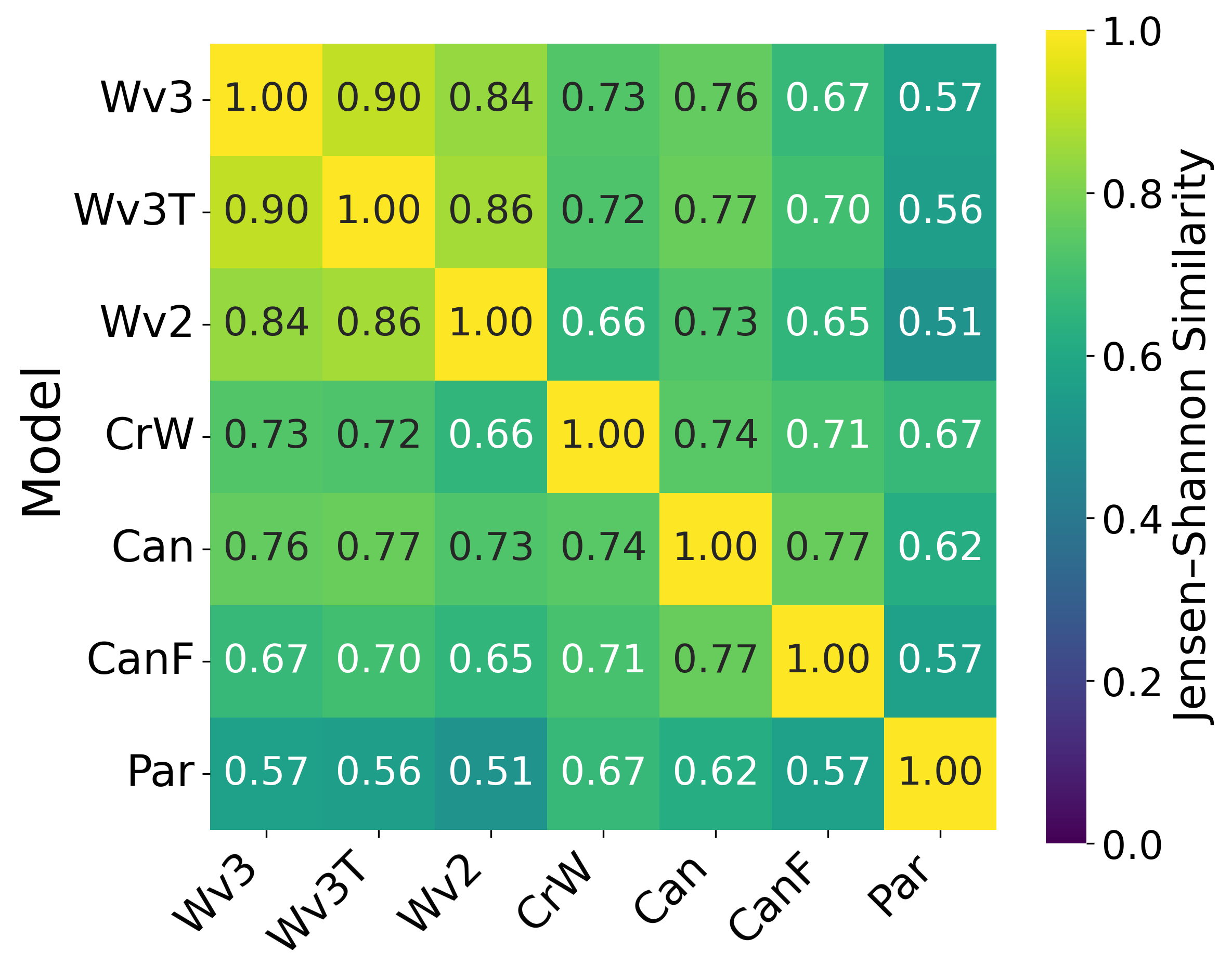}}
\end{minipage}
\vspace{-20pt}
\caption{Model hallucination distribution comparison.}
\label{fig:similarity}
\vspace{-5pt}
\end{figure}

\begin{figure}[!tb]
\vspace{-5pt}

\begin{minipage}[b]{1.0\linewidth}
 \centering
 \centerline{\includegraphics[width=8cm]{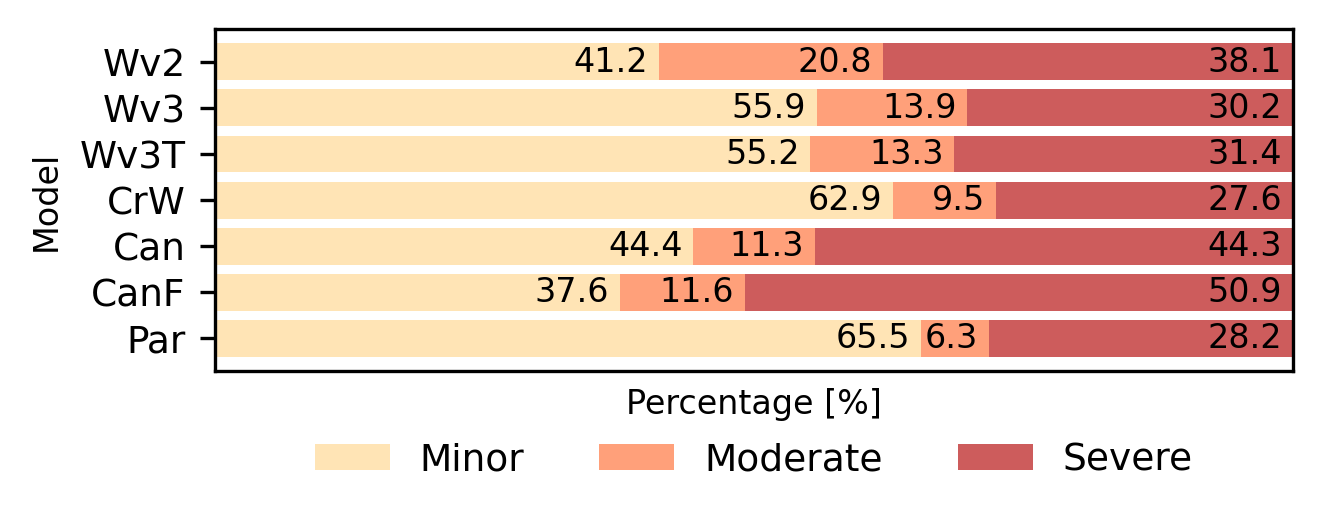}}
\end{minipage}

\vspace{-13pt}

\caption{Hallucination severity categories per ASR model.}
\label{fig:severity}
\vspace{-15pt}

\end{figure}

Thirteen phrases appeared in the lists of all models (\textit{you, okay, thank you, good, the, thank, but, that, yeah, and, ahead, question, you know}) and five appeared in all but one (\textit{yes, it, mr, much, so}). When taking only the top 10 hallucinations from each model, the most common phrases were: \textit{you, thank you, okay, good, and, thank, thanks, yeah, yes, ahead, it}. 

Figure \ref{fig:similarity} shows a comparison of the distributions of hallucinated phrases using Jensen-Shannon similarity \cite{endres2003new}. Hallucinations that appear only once are filtered out because they are often longer permutations of common hallucinated phrases. OpenAI Whispers (Wv3, Wv2, Wv3T) form a group with high similarity scores (\textgreater0.84) of hallucination text distributions. Par is a clear outlier with moderate similarity to most other models, including Can and CanF which are also Nemo models. Overall, this analysis demonstrates strong similarities between hallucinations generated by different ASR models.

\subsection{Severity of hallucinations estimated using a LLM}
\label{sec:llm}

Inspired by \cite{atwany2025lost}, we used GPT-4o mini\footnote{\url{https://developers.openai.com/api/docs/models/gpt-4o-mini}} to assess the severity of hallucinations by classifying each hallucination as minor (low-impact), moderate (partial obstruction/context shift) or severe (contradictory/meaning-altering). Each case was evaluated five times and the label was selected by majority vote. The agreement was high (Fleiss's kappa \textgreater 0.79) \cite{fleisskappa1971}. The severity proportions for each model are depicted in Fig.~\ref{fig:severity}, and the complete prompt is provided in the GitHub repository.

The severity of hallucination varies substantially across architectures. Par, CrW, and Wv3 mainly produce minor insertions, although more than 25\% of their hallucinations still severely distort meaning. In contrast, CanF, Can and Wv2 exhibit higher rates of severe errors (\textgreater38\%). Across all models, moderate hallucinations are least frequent, indicating that hallucination errors are either benign fillers or major semantic disruptions that contradict or alter meaning.

\section{Benchmarking Hallucination Detection}
\label{sec4}

\subsection{Detection with proxy metrics}

\begin{figure}[!t]

\begin{minipage}[b]{1.0\linewidth}
  \centering
  \centerline{\includegraphics[width=8cm]{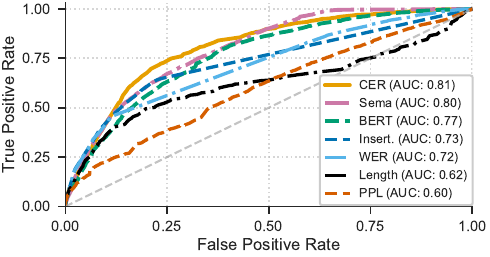}}
\end{minipage}
\vspace{-18pt}

\caption{ROC curves and AUC scores for various hallucination metrics across all ASR models.}
\label{fig:roc_curve}
\vspace{-15pt}
\end{figure}

Inspired by \cite{frieske2024hallucinations}, we demonstrate how to use HALAS to benchmark 7 text proxy metrics against human ground-truth hallucination labels for all 7 ASR models. The considered metrics might be categorized into three groups:
standard ASR structural metrics (WER, Character Error Rate (CER), Insertion Rate~\cite{levenshtein1966binary} and Length Ratio \cite{length_ratio}), semantic similarity metrics (BERTScore~\cite{bert} and SeMaScore~\cite{Sasindran_2024}) and language model features (GPT-2 Perplexity (PPL)~\cite{radford2019language}). Using metric values as direct predictors of hallucinations, we calculated the Receiver Operating Characteristic (ROC) curves and the Area Under the Curve (AUC)~\cite{fawcett2006roc} for all ASR predictions in HALAS. %
Figure \ref{fig:roc_curve} shows the results across all ASR models. 
The shapes of the ROC curves and the AUC in the range of 0.7-0.8 indicate that most of the structural and semantic metrics perform similarly, with CER and SeMaScore slightly outperforming the others, suggesting that combining those categories might result in improved performance. The lower efficacy of Perplexity and Length Ratio (AUC $\leq$ 0.62) confirms that hallucinations annotated in HALAS are often fluent and concise, indistinguishable from correct predictions via simple heuristics. 

Additionally, to assess the utility of the HALAS dataset, we test whether hallucination detection by combining these metrics generalizes across ASR models. For each model, we train an XGBoost classifier \cite{xgboost} on the 7 text proxy metrics on the HALAS training split using three regimes: only the target model's own training data (OWN), the training data of all other models with the target unseen (OTHER), and all training data (ALL). Averaged across the seven models, the classifier reaches a mean ROC-AUC of 0.829 (OWN), 0.828 (OTHER), and 0.835 (ALL). OTHER matching OWN indicates that HALAS can be used to train hallucination detectors for any ASR model, including ones absent from the dataset. Furthermore, AUC for ALL greater than for OWN shows that multi-model annotations of HALAS provide predictive value beyond any single architecture.

\subsection{Detection with SOTA methods}

In this section, we present the utility of HALAS by benchmarking two SOTA detection approaches for Whisper large v3: incorporating an LLM for comparing prediction to reference text~\cite{atwany2025lost} and based on the classification of decoder embeddings (DE) \cite{glazer2025transcriptionmechanisticinterpretabilityasr}. %
Results for GPT-4o mini and Gemini 2.0 Flash\footnote{\url{https://docs.cloud.google.com/gemini-enterprise-agent-platform/models/gemini/2-0-flash}} LLM models are also included. We replicate \cite{glazer2025transcriptionmechanisticinterpretabilityasr} exactly, passing the embedding from layer 21 of the Wv3 model decoder after generating End-of-Sequence token, into an utterance-level Logistic Regression (LR) classifier~\cite{logisticregression} (DE 21). We extended this approach to combinations of up to three layers, with the best results obtained for the concatenated outputs of layers 2, 13, and 23 (DE 2,13,23). LR weights and biases were fitted using the HALAS \textit{train} split. A full investigation of Whisper large v3 hallucination detection utilizing HALAS is presented in \cite{jasinski2026detection}.

\input{Tables/tab3}

Table \ref{tab:detectors} presents a comparison of detector performance on the HALAS \textit{test} split. Surprisingly, the reference-free DE detectors significantly outperformed both LLM-based models, despite the LLMs having direct access to the ground-truth reference texts. The two LLM models achieved very similar F1 scores, but through different flaws, with GPT-4o mini having low Precision and Flash 2 having low Recall. Proxy metric detectors perform strongly without decoder embedding access (ALL) and even without target model data (OTHER). The multi-layer DE detector outperformed the single-layer baseline, improving Recall and achieving the highest F1 score of 56.1\%.

\input{Tables/tab4}

Using the non-speech-augmented dataset from [4], where any ASR insertion relative to the reference is considered a hallucination, we evaluated the cross-domain generalization of the DE detectors trained on HALAS. As shown in Table \ref{tab:eos_detectors}, the DE detectors achieved a higher F1 score on these data than on the more difficult HALAS. This implies that HALAS can be used for the training of generalizable hallucination detectors and that its focus on difficult speech examples presents a challenging benchmark for detectors. %

\section{Conclusion}
\label{sec:conclusion}

In this work, we introduced HALAS, the first dataset of human-annotated ASR hallucinations on real-speech recordings. Our analysis shows that all seven evaluated state-of-the-art ASR models are prone to hallucinations, with generated errors exhibiting strong cross-model similarities and distributions heavily skewed toward a small set of phrases. We demonstrate the utility of the dataset by benchmarking common metrics alongside state-of-the-art detection methods, showing the difficulty of reliable hallucination detection, and establishing HALAS as a challenging benchmark for future research.

\section{Acknowledgments}

This research was supported by the National Science Centre, Poland under Grant 2021/42/E/ST7/00452, the National Centre for Research and Development, Poland under Grant INFOSTRATEG-IV/0029/2022, and by program "Excellence initiative – research university" for the AGH University of Krakow. We gratefully acknowledge Polish high-performance computing infrastructure PLGrid (HPC Centers: ACK Cyfronet AGH) for providing computer facilities and support within computational Grants  PLG/2025/018799 and PLG/2025/018238, and HumanSignal \cite{Label_Studio} for allowing use of their platform for annotation. %

\section{Generative AI Use Disclosure}
The authors used large language models (ChatGPT, Gemini, Claude) to assist with language editing. All experimental design, data processing, statistical analysis, and scientific conclusions were independently conducted and verified by the authors. The authors take full responsibility for the content of this manuscript.

\bibliographystyle{IEEEtran}
\bibliography{refs}

\end{document}

%% file: intro_mw.tex
\section{Introduction}
\label{sec:intro}

Although an impressive generalization of modern ASR models to various datasets and domains is achieved in zero-shot settings, the use of unreliable training data \cite{frieske2024hallucinations}, and the over-reliance on training patterns \cite{ye2024spurious} occasionally results in \textit{hallucinations}. Most researchers agree on the definition that ASR hallucinations are erroneous texts in the generated transcription that do not correspond to the words spoken in the audio input \cite{frieske2024hallucinations,ji2023survey,  baranski2025investigationwhisperasrhallucinations, carelesswhisper, atwany2025lost}. In this research, we follow a similar definition of hallucination: specific errors in ASR predictions, which vary from minor filler words to additions that change or contradict intent, or loopings, i.e. multiple repetitions of a word or phrase. The appearance of hallucinations in the ASR output can indicate a malfunctioning system and lower the user experience, or even lead to misinformation or misinterpretation, especially in a medical context~\cite{carelesswhisper, vishwanath2024faithfulness}.

The very recent methods for the detection of hallucinations might be divided into reference-based methods, which require ground truth text, and non-reference approaches. The first group utilizes reference text to validate whether the ASR output prediction contains hallucinations, e.g., by comparing the output prediction to the reference text using LLMs \cite{atwany2025lost}. Non-reference approaches involve detection based on a lookup-list of the most frequent phrases generated based on non-speech audio input~\cite{baranski2025investigationwhisperasrhallucinations} or binary classification of an embedding from a specific decoder layer \cite{glazer2025transcriptionmechanisticinterpretabilityasr}. Detection based on thresholding of reference-based or non-reference-based proxy metrics such as Word Error Rate (WER) and Perplexity \cite{frieske2024hallucinations} might be considered as a mixed approach.
Mitigation of hallucinations from ASR has been explored through pre-processing of audio input using Voice Activity Detection (VAD)~\cite{whisperx}, fine-tuning of either the weights of the ASR model~\cite{wagner2024crisperwhisperaccuratetimestampsverbatim} or its self-attention heads~\cite{wang2025calmwhisperreducewhisperhallucination} using noisy recordings devoid of speech, for which the model should return empty transcriptions.
The assessment of all the above-mentioned methods has been performed using audio datasets of non-speech or artificially corrupted speech, limiting their applicability in practice. To our knowledge, the occurrence of ASR hallucinations for unprocessed real speech has not been investigated so far. %

In this paper, we aim to fill this gap and introduce HALAS - \textbf{H}allucination \textbf{A}nnotations for \textbf{L}arge-scale \textbf{A}SR \textbf{S}ystems - the first publicly available dataset of human-labeled hallucinations from seven state-of-the-art ASR models on recordings from the Earnings 22 dataset \cite{earnings22}. First, we show the detailed description of dataset creation, including human annotation. Next, in the qualitative and quantitative analysis, we scrutinize the annotated ASR outputs, and show that all models indeed hallucinate on real data with the phrases of overlapping distributions. The investigation conducted with LLM indicates that a substantial amount of hallucination is severe regardless of the model. Finally, with an established HALAS benchmark that encompasses carefully prepared train and test splits, we evaluate hallucination detection based on multiple proxy metrics and SOTA methods, showing the challenging character of the problem studied using data with natural speech. We also show that a multi-layer extension of a decoder-embedding detector, trained on HALAS, improves upon the baseline for both in-domain and out-of-domain data. All in all, the contributions of the paper are (i) novel HALAS dataset, (ii) the first analysis of hallucinations of SOTA ASR models on real data, and (iii) the proposed benchmark that allows for evaluation of detection methods against naturally occurring, human-annotated hallucinations.

%% file: Tables/tab_HalasSplits.tex
\begin{table}[t]
    \centering
    \vspace{-10pt}
        \caption{Number of files, 5th and 95th file duration percentiles, and hallucination rates (HR) in the splits of HALAS dataset. }
        \vspace{-5pt}
    \begin{tabular}{l|cccc}
        \toprule
        Split & \# & $P_5$ [s] &$P_{95}$ [s] & HR [\%] \\
        \midrule
        \textit{train} & 2866 & 0.26 & 8.02 & 33.6\\
        \textit{test} & 745 & 1.22 & 12.15 & 22.6\\
        \bottomrule
    \end{tabular}

    \label{tab:benchmarks_splits}
    \vspace{-10pt}
\end{table}

%% file: Tables/tab1.tex
\begin{table*}[!t]
\caption{Summary of annotation results for each model, including annotated error count, hallucination and looping rates, WER, and the percentage of utterances contained within the top 10 and top 30 most commonly occurring phrases.}
\vspace{-6pt}
    \centering
    \begin{tabular}{@{}l|ccccccccc|c@{}}
    \toprule
    Model & Wv2 & Wv3 & Wv3T & CrW & Can & CanF & Par & Average\\\midrule
    Annotated Errors {[}\#{]} & 1518 & 858  & 1060  & 772  & 1213 & 1099  & 1217 & 1105 \\ 
    Hallucination Rate {[}\%{]} & 43.8 & 23.8 & 29.4 & 21.4 & 33.6 & 30.4 & 33.7 & 30.9\\ 
    Looping Rate {[}\%{]}       & 1.3  & 0.5  & 0.9  & 0.6  & 0.9  & 3.1  & 0.2 & 1.1\\ 
    \midrule
    WER {[}\%{]}       & 74.16  & 31.46  & 37.33  & 41.10 & 48.48  & 110.5  & 42.91 & 55.13\\
    \midrule
    Top 10 / 30 [\%] & 53 / 68 & 59 / 79 & 66 / 84 & 48 / 75 & 44 / 67 & 34 / 58 & 83 / 93 & 55 / 75\\
    \bottomrule
    \end{tabular}
    \label{tab:dataset}
    \vspace{-5pt}
\end{table*}

%% file: Tables/tab3.tex
\begin{table}[!t]
\caption{Results [\%] of all evaluated detectors on Wv3 HALAS.}
\centering
\vspace{-5pt}
\begin{tabular}{@{}l|cccc@{}}
\toprule
Detector & Acc. & Prec. & Rec.  & F1  \\\midrule
GPT-4o mini                   & 71.7 & 30.1 & 62.6 & 40.7 \\
Gemini 2.0 Flash                   & 84.5 & 50.0 & 35.7 & 41.6 \\
DE 21 & \textbf{87.1} & \textbf{57.8} & 49.1 & 53.1 \\
\midrule
OTHER                   & 83.3 & 47.6 & 60.9 & 53.4 \\
ALL                   & 83.7 & 48.7 & \textbf{64.3} & 55.4 \\
DE 2,13,23 & 86.4 & 53.9 & 58.5 & \textbf{56.1} \\
\bottomrule
\end{tabular}
\label{tab:detectors}
\vspace{-5pt}
\end{table}

%% file: Tables/tab4.tex
\begin{table}[!t]
\caption{Results [\%] of DE detectors on out-of-domain data~\cite{baranski2025investigationwhisperasrhallucinations}.}
\centering
\vspace{-5pt}
\begin{tabular}{@{}l|cccc@{}}
\toprule
Detector & Acc. & Prec. & Rec. & F1 \\\midrule
DE 21 & 72.5 & 67.6 & \textbf{77.9} & 72.4 \\
DE 2,13,23 & \textbf{80.0} & \textbf{81.6} & 73.3 & \textbf{77.3} \\
\bottomrule
\end{tabular}
\label{tab:eos_detectors}
\vspace{-15pt}
\end{table}

%% file: refs.bib
@inproceedings{jasinski2026detection,
  author    = {Jasi{\'n}ski, Jan and Bara{\'n}ski, Mateusz and Bartolewska, Julitta and Witkowski, Marcin and Kowalczyk, Konrad},
  title     = {{From Text Metrics to Model Internals: A Study of {Whisper} {ASR} Hallucination Detection}},
  booktitle = {Proceedings of Interspeech},
  year      = {2026}
}

@article{logisticregression,
author = {Peng, Joanne and Lee, Kuk and Ingersoll, Gary},
year = {2002},
title = {{An Introduction to Logistic Regression Analysis and Reporting}},
journal = {Journal of Educational Research},
volume = {96},
doi = {10.1080/00220670209598786}

}

@misc{Label_Studio,
title={{Label Studio}: Data labeling software},
url={https://github.com/HumanSignal/label-studio},
note={open source software available from \url{https://github.com/HumanSignal/label-studio}},
author={
Maxim Tkachenko and
Mikhail Malyuk and
Andrey Holmanyuk and
Nikolai Liubimov},
year={2020-2025},
}

@inproceedings{xgboost,
   title={{XGBoost: A Scalable Tree Boosting System}},
   DOI={10.1145/2939672.2939785},
   booktitle={Proceedings of the 22nd ACM SIGKDD International Conference on Knowledge Discovery and Data Mining},
   author={Chen, Tianqi and Guestrin, Carlos},
   year={2016},
}

@article{ye2024spurious,
  title={{Spurious Correlations in Machine Learning: A Survey}},
  author={Ye, Wenqian and Zheng, Guangtao and Cao, Xu and Ma, Yunsheng and others},
  journal={arXiv preprint arXiv:2402.12715},
  year={2024}
}

@InProceedings{radford2023robust,
  title = 	 {{Robust Speech Recognition via Large-Scale Weak Supervision}},
  author =       {Radford, Alec and Kim, Jong Wook and Xu, Tao and Brockman, Greg and others},
  booktitle = 	 {{Proceedings of the 40th International Conference on Machine Learning}},
  year = 	 {2023},
  volume = 	 {202},

}

@article{ji2023survey,
  title={{Survey of Hallucination in Natural Language Generation}},
  author={Ji, Ziwei and Lee, Nayeon and Frieske, Rita and Yu, Tiezheng and others},
  journal={ACM Computing Surveys},
  year={2023},
  volume = {55},
  number = {12},
  publisher={ACM New York, NY},
  DOI={10.1145/3571730},
}

@article{frieske2024hallucinations,
  title={{Hallucinations in Neural Automatic Speech Recognition: Identifying Errors and Hallucinatory Models}},
  author={Frieske, Rita and Shi, Bertram E},
  journal={arXiv preprint arXiv:2401.01572},
  year={2024}
}

@inproceedings{baranski2025investigationwhisperasrhallucinations,
  title={{Investigation of Whisper {ASR} Hallucinations Induced by Non-Speech Audio}},
  author={Bara{\'n}ski, Mateusz and Jasi{\'n}ski, Jan and Bartolewska, Julitta and Kacprzak, Stanis{\l}aw and others},
  booktitle={Proceedings of IEEE International Conference on Acoustics, Speech and Signal Processing (ICASSP)},
  year={2025}
}

@inproceedings{carelesswhisper,
  title={{Careless {W}hisper: Speech-to-Text Hallucination Harms}},
  author={Koenecke, Allison and Choi, Anna Seo Gyeong and Mei, Katelyn X and Schellmann, Hilke and Sloane, Mona},
  booktitle={{Proceedings of the ACM Conference on Fairness, Accountability, and Transparency}},
  year={2024}
}

@inproceedings{whisperx,
  title={{WhisperX: Time-Accurate Speech Transcription of Long-Form Audio}},
  author={Bain, Max and Huh, Jaesung and Han, Tengda and Zisserman, Andrew},
  booktitle={{Proceedings of Interspeech}},
  year={2023}
}

@inproceedings{wang2025calmwhisperreducewhisperhallucination,
      title={{Calm-Whisper: Reduce Whisper Hallucination On Non-Speech By Calming Crazy Heads Down}}, 
      author={Yingzhi Wang and Anas Alhmoud and Saad Alsahly and Muhammad Alqurishi and others},
      booktitle={{Proceedings of Interspeech}},
      year={2025}
}

@article{glazer2025transcriptionmechanisticinterpretabilityasr,
      title={{Beyond Transcription: Mechanistic Interpretability in ASR}}, 
      author={Neta Glazer and Yael Segal-Feldman and Hilit Segev and Aviv Shamsian and others},
      journal={arXiv preprint arXiv:2508.15882},
      year={2025}
}

@inproceedings{puvvada2024moreaccuratespeechrecognition,
  title     = {{Less is More: Accurate Speech Recognition \& Translation without Web-Scale Data}},
  author    = {Krishna C. Puvvada and Piotr Żelasko and He Huang and Oleksii Hrinchuk and Nithin Rao Koluguri and others},
  year      = {2024},
  booktitle = {{Proceedings of Interspeech}},
  doi       = {10.21437/Interspeech.2024-2294},
  issn      = {2958-1796},
}

@article{earnings22,
      title={{Earnings-22: A Practical Benchmark for Accents in the Wild}}, 
      author={Miguel Del Rio and Peter Ha and Quinten McNamara and Corey Miller and Shipra Chandra},
      journal={arXiv preprint arXiv:2203.15591},
      year={2022}
}

@article{zelasko2025traininginferenceefficiencyencoderdecoder,
      title={{Training and Inference Efficiency of Encoder-Decoder Speech Models}}, 
      author={Piotr Żelasko and Kunal Dhawan and Daniel Galvez and Krishna C. Puvvada and others},
      journal={arXiv preprint arXiv:2503.05931},
      year={2025}
}

@InProceedings{xu2023efficientsequencetransductionjointly,
  title = 	 {{Efficient Sequence Transduction by Jointly Predicting Tokens and Durations}},
  author =       {Xu, Hainan and Jia, Fei and Majumdar, Somshubra and Huang, He and others},
  booktitle = 	 {Proceedings of the 40th International Conference on Machine Learning},
  year = 	 {2023},
}

@article{fleisskappa1971,
  title={Measuring nominal scale agreement among many raters},
  author={Joseph L. Fleiss},
  journal={Psychological Bulletin},
  year={1971},
  volume = {76},
  number = {5}
}

@article{endres2003new,
  title={A new metric for probability distributions},
  author={Endres, Dominik Maria and Schindelin, Johannes E},
  journal={IEEE Transactions on Information Theory},
  year={2003},
  volume={49},
  number={7},
  publisher={IEEE}
}

@inproceedings{wagner2024crisperwhisperaccuratetimestampsverbatim,
  title={{CrisperWhisper: Accurate Timestamps on Verbatim Speech Transcriptions}},
  author={Wagner, Laurin and Thallinger, Bernhad and Zusag, Mario},
  booktitle={{Proceedings of Interspeech}},
  year={2024}
}

@article{atwany2025lost,
  title={{Lost in Transcription, Found in Distribution Shift: Demystifying Hallucination in Speech Foundation Models}},
  author={Atwany, Hanin and Waheed, Abdul and Singh, Rita and Choudhury, Monojit and others},
  journal={arXiv preprint arXiv:2502.12414},
  year={2025}
}

@article{cohen1960coefficient,
  title={{A Coefficient of Agreement for Nominal Scales}},
  author={Cohen, Jacob},
  journal={Educational and Psychological Measurement},
  year={1960},
  volume = 20,
  number = 1,
  publisher={Sage Publications Sage CA: Thousand Oaks, CA}
}

@article{srivastav2025openasrleaderboardreproducible,
      title={{Open ASR Leaderboard: Towards Reproducible and Transparent Multilingual and Long-Form Speech Recognition Evaluation}}, 
      author={Vaibhav Srivastav and Steven Zheng and Eric Bezzam and Eustache Le Bihan and others},
      year={2025},
      journal={arXiv preprint arXiv:2510.06961},
}

@inproceedings{levenshtein1966binary,
  title={Binary codes capable of correcting deletions, insertions, and reversals},
  author={Levenshtein, Vladimir I},
  booktitle={Soviet physics doklady},
  volume={10},
  number={8},
  year={1966}
}

@article{bert,
      title={{BERTScore: Evaluating Text Generation with {BERT}}}, 
      author={Tianyi Zhang and Varsha Kishore and Felix Wu and Kilian Q. Weinberger and others},
      year={2020},
      journal={arXiv preprint arXiv:1904.09675}, 
}

@inproceedings{Sasindran_2024,
   title={{SeMaScore: A new evaluation metric for automatic speech recognition tasks}},
   DOI={10.21437/interspeech.2024-2033},
   booktitle={{Proceedings of Interspeech}},
   author={Sasindran, Zitha and Yelchuri, Harsha and Prabhakar, T. V.},
   year={2024},
   collection={interspeech_2024} 
}

@article{radford2019language,
  author = {Radford, Alec and Wu, Jeffrey and Child, Rewon and Luan, David and others},
  journal = {OpenAI},
  title = {{Language Models are Unsupervised Multitask Learners}},
  note = {Accessed: 2024-11-15},
  url = {https://cdn.openai.com/better-language-models/language_models_are_unsupervised_multitask_learners.pdf},
  year = 2019
}

@article{fawcett2006roc,
  title={{An introduction to ROC analysis}},
  author={Fawcett, Tom},
  journal={Pattern Recognition Letters},
  volume = {27},
  number = {8},
  year={2006}
}

@inproceedings{vishwanath2024faithfulness,
  title={{Faithfulness Hallucination Detection in Healthcare {AI}}},
  author={Vishwanath, Prathiksha Rumale and Tiwari, Simran and Naik, Tejas Ganesh and Gupta, Sahil and others},
  booktitle={Artificial Intelligence and Data Science for Healthcare: Bridging Data-Centric AI and People-Centric Healthcare},
  year={2024}
}

@article{length_ratio,
  author  = {Angela Fan and Shruti Bhosale and Holger Schwenk and Zhiyi Ma and Ahmed El-Kishky and others},
  title   = {{Beyond English-Centric Multilingual Machine Translation}},
  journal = {{Journal of Machine Learning Research}},
  year    = {2021},
volume = {22},
  number = {1},
}
